\documentclass{PoS}

\usepackage[utf8]{inputenc}
\usepackage{fixltx2e}
\usepackage{mathtools}
\usepackage{slashed}
\usepackage{bbm}
\usepackage{mleftright}
\usepackage{nicefrac}
\usepackage[caption=false]{subfig}
\usepackage[
	style=numeric-comp,
	backend = bibtex,
	bibencoding = ascii,	
	sorting=none,
	firstinits=true,
	isbn=false,
	url=false,
	doi=false,
	maxbibnames = 99
	]{biblatex}
\DeclareFieldFormat
  [article,inbook,incollection,inproceedings,patent,thesis,unpublished,suppbook,suppcollection,suppperiodical]
  {title}{\mkbibemph{#1}}
\DeclareFieldFormat{eprint:arxiv}{[#1]}

\renewbibmacro{in:}{}

\newcommand{\dx}{\mathrm{d}}

\newcommand{\ii}{\mathrm{i}}
\newcommand{\Nf}{{N_\mathrm{f}}}
\newcommand{\Nfc}{{N_\mathrm{f}^\mathrm{cr}}}
\newcommand{\Nftwo}{N_\mathrm{f,irr}}
\newcommand{\lGN}{\lambda_\mathrm{GN}}
\newcommand{\lTh}{\lambda_\mathrm{Th}}

\newcommand{\ex}[1]{\mathrm{e}^{#1}}
\newcommand{\srm}[1]{_\mathrm{#1}}

\DeclareMathOperator{\im}{Im}

\title{Four-Fermion Theories with Exact Chiral Symmetry in Three Dimensions}

\ShortTitle{Four-Fermion Theories with Exact Chiral Symmetry in Three Dimensions}

\author{\speaker{Daniel Schmidt}\\%
      Theoretisch-Physikalisches Institut, Friedrich-Schiller-Universit\"at Jena, 07743 Jena, Germany\\
        E-mail: \email{d.schmidt@uni-jena.de}}

\author{Björn Wellegehausen\\
       Institut f\"ur Theoretische Physik, Justus-Liebig-Universit\"at Giessen, 35392 Giessen, Germany and\\
      Theoretisch-Physikalisches Institut, Friedrich-Schiller-Universit\"at Jena, 07743 Jena, Germany
       E-mail: \email{bjoern.wellegehausen@uni-jena.de}}

\author{Andreas Wipf\\
       Theoretisch-Physikalisches Institut, Friedrich-Schiller-Universit\"at Jena, 07743 Jena, Germany\\
       E-mail: \email{wipf@tpi.uni-jena.de}}  

\abstract{
We investigate a class of four-fermion theories which includes well-known models like the Gross-Neveu model and the Thirring model.
In three spacetime dimensions, they are used to model interesting solid state systems like high temperature superconductors and graphene.
Additionally, they serve as toy models to study chiral symmetry breaking (CSB).

For any number of fermion flavours the Gross-Neveu model has a broken and a symmetric phase, while the existence of a broken phase in the Thirring model depends on the number of flavours.
The critical number of fermion flavours beyond which there exists no CSB is still subject of ongoing discussions.
Using SLAC fermions we simulate the Thirring model with exact chiral symmetry.
To obtain a chiral condensate one can introduce a symmetry-breaking mass term and carefully study the limits of infinite lattice and zero-mass.
So far, we did not see CSB within this approach for the Thirring model with $2$ or more (reducible) flavours.

The talk presents alternative approaches to investigate these findings.
We employ certain Fierz identities to map the Thirring model into equivalent four-fermion models, for which the chiral condensate does not seem to vanish. 
In the new formulations based on reshuffled degrees of freedom we find a sign problem (which is not present in the original formulation). 
For this reason we developed an algorithm similar to fermion bags, which may solve this problem.
As a further approach, we embed the multi-flavour Thirring model in a larger class of four-fermion theories to study the chiral symmetry and its breaking in a wider context.
}

\FullConference{34th annual International Symposium on Lattice Field Theory\\
         24-30 July 2016\\
         University of Southampton, UK}

\begin{document}

\section{Introduction}
We investigate a class of $\Nf$ fermion-flavour field theories with quartic interactions.
The Euclidean Lagrangian has the general form
  \begin{equation}
    \mathcal{L}=\bar\psi_j \left( \slashed{\partial} + m \right)\psi_j
    + \sum_\alpha \frac{g^2_\alpha}{2\Nf} \left(\bar\psi_j \Gamma_\alpha \psi_j\right)^2 \qquad j=1,\dots,\Nf,
  \end{equation}
where $\Gamma_\alpha$ could be any (antisymmetrised) product of $\gamma$-matrices or the identity.
Notable examples are the Thirring model ($\Gamma_\alpha=\gamma_\mu$) \cite{Thirring1958}, which can be solved analytically in $2$ dimensions, the Nambu-Jona-Lassinio model introduced to study dynamical mass generation in $4$ dimensions ($\Gamma_1=\mathbbm{1}$ and $\Gamma_2=\ii\gamma_5)$ \cite{Nambu1961}, and the $2$-dimensional Gross-Neveu model ($\Gamma=\mathbbm{1}$), as toy-model to study asymptotic freedom and CSB \cite{Gross1974}.

We are mainly interested in 3-dimensional versions of the massless Thirring model, because it is similar to QED$_3$ and may have applications in condensed matter systems like superconductors and graphene (see for example \cite{Hands2008,Herbut2009}).
Having these applications in mind many authors considered the model with a reducible representation of the Clifford algebra, where the $\gamma_\mu$  for $\mu = 1, 2, 3$ are the usual matrices known from 4 dimensions acting on $4$-component spinors.
We shall follow this habit in Section \ref{s:2ff} where we use a $4$-dimensional reducible representation as well.
In sections~\ref{s:fierz} and~\ref{s:fb} we instead use an irreducible  $2$-dimensional representation of the Clifford algebra.

The massless Thirring model with $\Nf$ reducible flavours has an  enlarged $U(\Nf,\Nf)$ symmetry generated by the matrices $\{\mathbbm{1},\ii\gamma_4,\ii\gamma_5,\ii\gamma_4\gamma_5\}$ tensored with anti-hermitian matrices acting on the flavour indices. 
Here $\gamma_1,\dots,\gamma_4$ and $\gamma_5=\gamma_1\gamma_2\gamma_3\gamma_4$ are $\gamma$-matrices in four (Euclidean) dimensions. 
A mass term or  chiral condensate breaks the symmetry explicitly or spontaneously to $U(\Nf)\otimes U(\Nf)$.
In an irreducible representation with $2$-component spinors the model has $\Nftwo=2\Nf$ flavours. 
The Thirring model with one irreducible flavour (which cannot be written in terms of reducible spinors) is equivalent to the Gross-Neveu model with one irreducible flavour which shows CSB \cite{Reisz1998}. 
On the other hand, since a broken phase is absent in the large-$\Nf$ expansion of the Thirring model there must exist a critical flavour  number $\Nfc>\nicefrac{1}{2}$ (or equivalently $\Nftwo>1$) such that symmetry breaking occurs only for $\Nf<\Nfc$.
There have been many different predictions for $\Nfc$ based on Schwinger-Dyson equations \cite{Itoh1995,Gomes1991,Hong1994}, $\nicefrac{1}{\Nf}$-expansion \cite{Kondo1995}, functional renormalization group \cite{Janssen2012-2} and lattice simulations \cite{DelDebbio1997-1,Kim1996,Christofi2007,DelDebbio1999,Hands1999}.
Most of these studies find $\Nfc$ between 2 and 7, but there is no agreement on a concrete value.

Our goal is to improve the lattice results for $\Nfc$. 
This is desirable, since all previous attempts used staggered fermions which do not admit the full chiral symmetry. 
Hence it is not clear,  whether in the continuum limit one ends up within the universality class of the Thirring model.
We presented first results with SLAC fermions in \cite{Schmidt2015}.
This approach allows simulations with exact chiral symmetry, but for  the reducible model we did not see a non-vanishing chiral condensate.
Further investigations with explicit symmetry breaking terms (global fields and a mass term) showed the same results after the trigger was removed. Although a reliable estimate was not possible,  these simulations indicate a small value of $\Nfc \approx 2$.
Remarkably, this is in agreement with recent simulations with domain wall fermions \cite{Hands2016}, where no sign of symmetry breaking was found at $\Nf=2$.

The remaining sections each shortly present a different route to shed further light on the still unsatisfactory situation regarding chiral symmetry and its breaking in the Thirring model.

\section{Two Four-Fermion Interactions}
\label{s:2ff}
One approach to study CSB is to enlarge the theory space.
This is motivated by observations within the framework of functional renormalization group, where the fixed point governing the critical behaviour of the Thirring model is found outside of the class of Thirring models \cite{Janssen2012-2}.
In addition, the enlarged theory space contains the Gross-Neveu model, the expected phase structure of which is well-reproduced with simulations  based on SLAC fermions.

\begin{figure}[htb]
  \centering
  \input{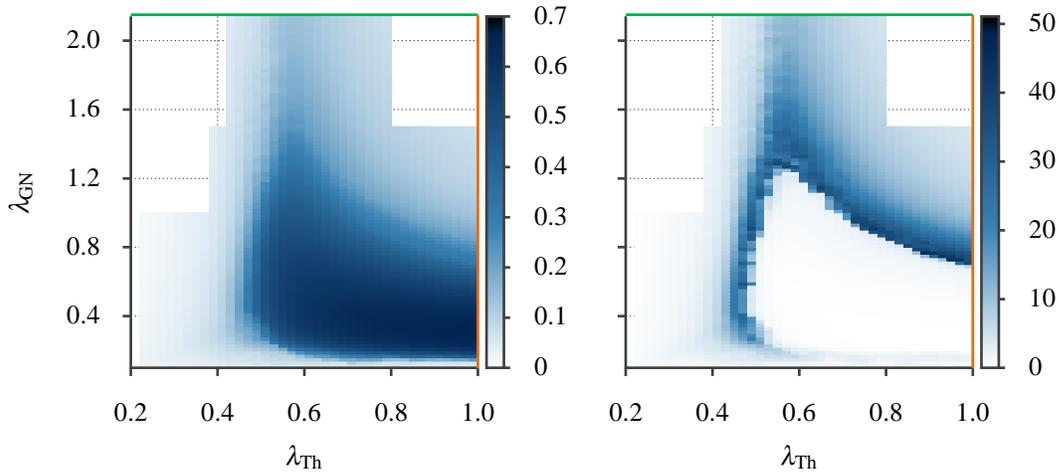}
  \caption{Chiral condensate (left) and susceptibility (right) of the coupled Gross-Neveu and Thirring model for $\Nf=1$ and lattice size $8\times7\times7$. The right (orange) edge of each plot is in the direction of the pure Gross-Neveu model, while the top (green) edge is in the direction of the pure Thirring model.}
  \label{f:gnth}
\end{figure}

First results for the coupled $\Nf=1$ model with Lagrangian 
\begin{equation}
  \mathcal{L}=\bar\psi_j \slashed{\partial}\psi_j
    + \frac{1}{4\Nf\lGN} \left(\bar\psi_j \psi_j\right)^2
    + \frac{1}{4\Nf\lTh} \left(\bar\psi_j \gamma_\mu \psi_j\right)^2
\end{equation}
can be found in Figure~\ref{f:gnth}.
There is a large region with non-vanishing chiral condensate, which bends towards the pure Thirring model with $\lambda\srm{GN}\rightarrow\infty$.
Although the maximal value of the condensate is decreasing with increasing $\lambda\srm{GN}$, there is some trace of CSB left for finite $\lambda\srm{GN}$.
The susceptibility on the right hand side of Figure~\ref{f:gnth} shows a clear peak at the physical phase transition of the Gross-Neveu model and another peak at smaller $\lTh$, where the condensate decreases again.
As in our previously reported results with a global symmetry-breaking field~\cite{Schmidt2015}, we interpret this as a transition to a lattice artefact phase.
Again similar to our previous findings, these two transitions merge when  approaching the pure Thirring model. 
CSB is possible for $\Nf=1$, but the existence of the lattice artefact phase  does not allow for a final conclusion.

\begin{figure}[htbp]
  \centering
    \input{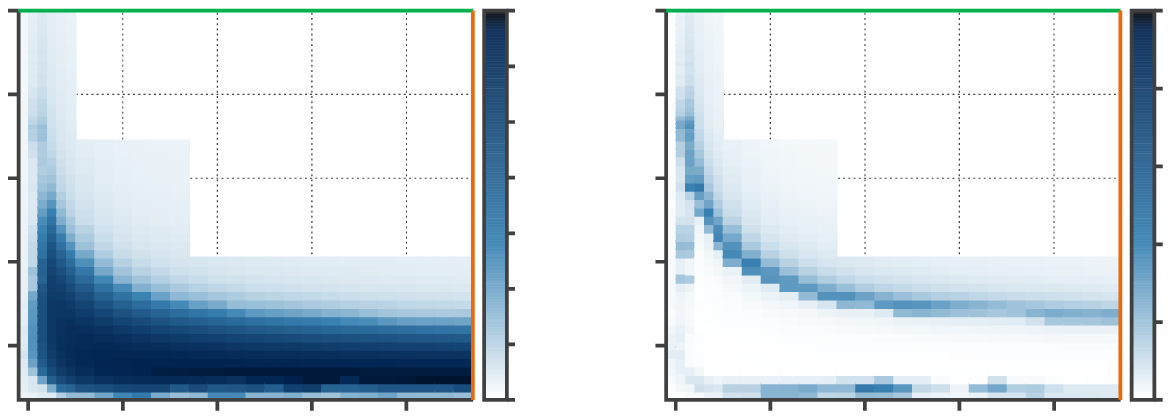}
  
    \input{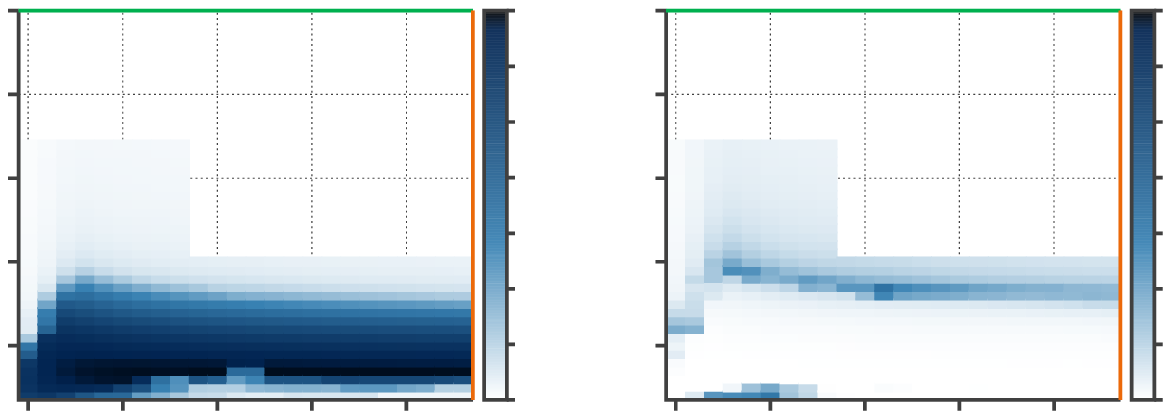}
  
    \input{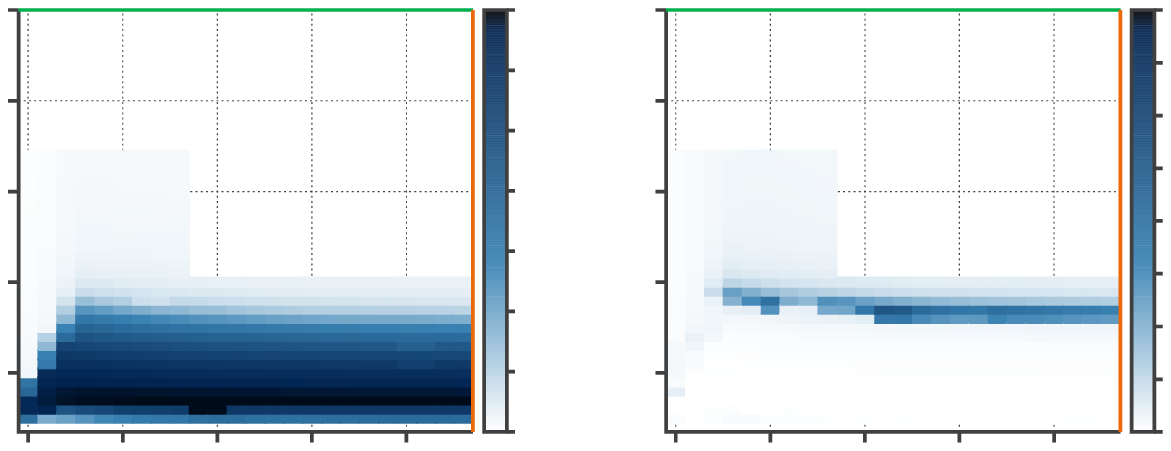}
  \caption{Chiral condensate (left) and susceptibility (right) of the Gross-Neveu model with additional $\gamma_{45}$-term (see main body of the text) for $\Nf=1, 2, 3$ and lattice size   $12\times11\times11$. The right (orange) edge of the plot is in the direction of the pure Gross-Neveu model, while the top (green) edge is in the direction of the pure $\gamma_{45}$-model.}
  \label{f:gn45}
\end{figure}
Motivated by \cite{Gehring2015} we also studied a model with interaction $\pm\frac{1}{4\Nf\lambda_{45}}\left(\bar\psi \ii\gamma_{4}\gamma_{5}\psi\right)^2$, which shares the full $U(\Nf,\Nf)$ symmetry  with the Thirring model.
The model with negative sign should behave like a Gross-Neveu model, but shows a sign problem and thus is not accessible through conventional  lattice simulations.
Hence we will present here results for the model with positive sign.
Then there is no sign problem and the model behaves similar to the Thirring model.
There is no direct evidence for CSB in the single $\gamma_{45}$-model, but the $\Nf=1$ phase space for the model coupled to the Gross-Neveu model in Figure~\ref{f:gn45} looks similar to Figure~\ref{f:gnth}.
For this simpler model, we also simulated $\Nf=2,3$ and $4$.
Figure~\ref{f:gn45} shows a slight bending of the phase transition line towards the pure $\gamma_{45}$-model axis, while there is no evidence for this for $\Nf=3$.
$\Nf=4$ is not shown here, since the plot looks quite similar to $\Nf=3$.
While this is only a very indirect signal and further investigations of the coupled Thirring model for larger flavour numbers and lattice sizes are needed, it again points to a rather small critical flavour number of the Thirring model.
\section{Fierz transformations}
\label{s:fierz}
\begin{figure}[hbt]
  \centering
  \begin{minipage}{.475\linewidth}
    \input{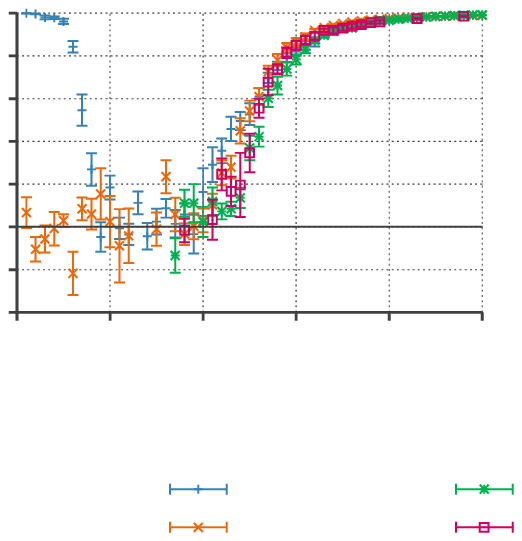}
    \caption{The weight factor $\left<w\right>=\left<\ex{-\ii\im S}\right>$ for different numbers of flavours in the Fierz version \protect\eqref{e:fierz_lagrangian} of the Thirring model with $6\times5\times5$ lattice points.}
    \label{f:weights}
  \end{minipage}%
  \hfill%
  \begin{minipage}{.475\linewidth}
    \input{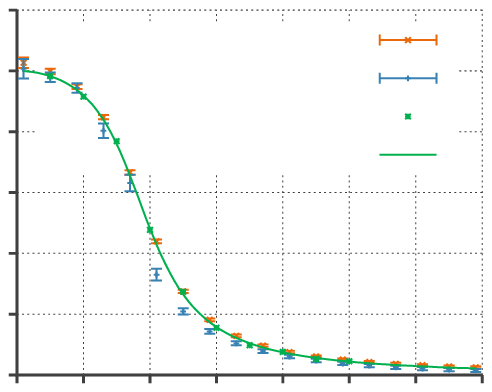}
    \caption{Average value of the $k$-field of the new fermion bag approach with $\Nftwo = 1$ from analytical computations and Metropolis simulations on a $2\times3\times3$ lattice, compared with results from ordinary rHMC simulations.}
    \label{f:fermionbag}
   \end{minipage}
\end{figure}

Another approach to obtain a directly accessible condensate for the Thirring model is to use Fierz identities, relating different kinds of four-fermion interactions with each other.
From now on, we use the irreducible representation of the Clifford algebra, since then  the identities are much simpler.
The two-component spinors are denoted by $\chi_a$ and the number of irreducible flavours is twice the number of reducible flavours: $a=1,\dots,2\Nf\coloneqq \Nftwo$.
For our purposes, the following identity is the most useful:
\begin{equation}
      \left(\bar\chi^a \sigma_\mu \chi^a\right)\left(\bar\chi^b \sigma^\mu \chi^b\right)
      = -\left(\bar\chi^a \chi^a\right)
	\left(\bar\chi^b \chi^b\right)
      -2\left(\bar\chi^a \chi^b\right)\left(\bar\chi^b \chi^a \right).
\end{equation}
After performing a Hubbard-Stratonovich transformation, we obtain the  equivalent Lagrangian
\begin{equation}
  \mathcal{L}\srm{Fierz}=\bar\chi_a
  \underbrace{\left[\left(\slashed\partial+\phi\right)\delta^{ab}+ T^{ab}\right]}_{=D^{ab}}
  \chi_b
  + \frac{\Nftwo}{4g^2}T_{ab}T^{ba}
  + \frac{\Nftwo}{2g^2}\phi^2,
  \label{e:fierz_lagrangian}
\end{equation}
where an auxiliary scalar field $\phi$ and a hermitian matrix field $T_{ab}$ were introduced.
The theory now contains $\Nf^2+1$ real scalar degrees of freedom, which makes simulations with large $\Nf$ expensive.
Even worse, the Dirac operator $D^{ab}$ is not anti-hermitian any more.
We do not expect a real and positive spectrum, contrary to the original  formulation of the irreducible Thirring model with even $\Nftwo$.
We performed simulations with an exact update algorithm and found indeed a  very strong sign problem. 
The action has a complex phase and, as shown in Figure~\ref{f:weights}, there is a region in coupling space, where its averages are very close to zero, such that  naive approaches like reweighting are not possible.

There are further possible rearrangements with Fierz identities, but they include even more degrees of freedom and none of them is free of the sign problem.
For example, there is an identity that allows for an anti-hermitian Dirac operator, such that the eigenvalues are purely imaginary.
Now the action is real, but the sign problem is even worse, because the sign of the action switches on nearly every update.
Thus, Fierz identities do not provide a new way to gain insight into the Thirring model, although the ``reverse'' application shows, that there are models with four-fermion interactions, where we can get rid of the sign problem by using these identities.

\section{Fermion Bag Approach}
\label{s:fb}
In order to solve the sign problem of the models after Fierz transformation, we are currently investigating an approach similar to the fermion bags first introduced in \cite{Chandrasekharan2010}.
Performing the integrals in the partition sum of the model given by \eqref{e:fierz_lagrangian} over the fermions and the additional scalar fields, we  get a dual formulation with a spin field $k_{xi}^{ab}\in\{0,1\}$ and a final  partition sum of
\begin{equation}
  Z(\lambda) \propto \sum_k (\lambda)^{-\frac{k}{2}} \det(\slashed\partial[k])2^{\tilde n^2_x}\prod_x f(n^1_x, n^2_x).
  \label{e:z_fb}
\end{equation}
$\slashed\partial[k]$ is the SLAC operator matrix, where rows and columns are deleted, when the corresponding value of $k_{xi}^{ab}=1$.
The values $n_x^1, n_x^2$ and $\tilde n_x^2$ count certain entries of the field and the function $f(a,b)$ is a combination of gamma- and confluent hypergeometric functions, which give an additional local weight.
Interestingly, this approach solves the sign problem for $\Nftwo=1$, present in both the original and the Fierz version of the irreducible Thirring model.
Metropolis simulations of \eqref{e:z_fb} work well and are found to agree on very small lattices with the full analytically computed partition sum (see Figure~\ref{f:fermionbag}).
Increasing the number of flavours poses new challenges and is still subject of ongoing research.

\acknowledgments
D.S. and B.W. were supported by the DFG Research Training Group 1523/2 ``Quantum and Gravitational Fields''. B.W. was supported by the Helmholtz International Center for FAIR within the LOEWE initiative of the State of Hesse. We thank Holger Gies and Lukas Janssen for insightful discussions.

\printbibliography

\end{document}